\documentclass[aps,pra,showpacs,twocolumn]{revtex4}
\usepackage{amsmath,graphicx,epsfig,color}

\newcommand{\ket}[1]{|#1\rangle}
\newcommand{\bra}[1]{\langle#1|}
\newcommand{\ketbra}[1]{\ket{#1}\bra{#1}}
\newcommand{\tr}{\text{tr}}
\newcommand{\ten}{\otimes}
\newcommand{\unit}{\mathbf{1}}

\newcommand{\half}{\frac{1}{2}}
\newcommand{\Slhv}{\mathcal{S}_\text{lhv}}

\newcommand{\eBellCH}{\langle\mathcal{B}_{CH}\rangle}
\newcommand{\eBellCHSH}{\langle\mathcal{B}_{CHSH}\rangle}

\begin{document}

\title{Better Bell Inequality Violation by Collective Measurements}

\author{Yeong-Cherng~Liang}
\email{ycliang@physics.uq.edu.au}

\author{Andrew~C.~Doherty}
\email{doherty@physics.uq.edu.au}
\affiliation{School of Physical Sciences, The University of
Queensland, Queensland 4072, Australia.}

\date{\today}

\begin{abstract}
The standard Bell inequality experiments test for violation of local
realism by repeatedly making local measurements on individual copies
of an entangled quantum state. Here we investigate the possibility
of increasing the violation of a Bell inequality by making
collective measurements. We show that the nonlocality of bipartite pure
entangled states, quantified by their maximal violation of the
Bell-Clauser-Horne inequality, can always be enhanced by collective
measurements, even without communication between the parties. For
mixed states we also show that collective measurements can increase
the violation of Bell inequalities, although numerical evidence
suggests that the phenomenon is not common as it is for pure states.
\end{abstract}

\maketitle

\section{Introduction}

It is one of the most remarkable features of quantum physics that
measurements on separated systems cannot always be described by
local realistic theories
~\cite{J.S.Bell:1964,CHSH:PRL:1969,CH:PRD:1974,N.Gisin:PLA:1991,N.Gisin:PLA:1992,S.Popescu:PLA:1992}.
Typically, this phenomenon is revealed by the violation of a Bell
inequality, which are constraints that have to be satisfied by any
local realistic description. Bell inequality violations have been
observed experimentally in various physical systems, such as
entangled photon pairs, as reviewed in~\cite{A.Aspect:Nature:1999} and entangled
$^9{\rm Be}^+$ ions~\cite{M.A.Rowe:Nature:1999}. For a background on
Bell inequalities readers are referred
to~\cite{R.F.Werner:QIC:2001}, and references therein.

Usually, experiments to test Bell inequalities involve making many
measurements on individual copies of the quantum system with the
system being prepared in the same way for each measurement. In this
paper, we consider a somewhat different scenario and ask if
quantum nonlocality can be enhanced by making joint local
measurements on multiple copies of the entangled state. We will use
the maximal Bell inequality violation of a quantum state $\rho$ as
our measure of nonlocality. Our interest is to determine if
$\rho^{\otimes N}$, when compared with $\rho$, can give rise to a
higher Bell inequality violation for some $N>1$.

A very similar problem was introduced by
Peres~\cite{A.Peres:PRA:1996} who considered Bell inequality
violations under collective measurements but allowed the
experimenters to make an auxiliary measurement on their systems and
postselect on both getting a specific outcome of their measurement.
Numerically, Peres showed that with collective measurements and
postselection~\cite{fn:Peres}, a large class of two-qubit states give rise to better
Bell inequality violation. Moreover, explicit examples were given to
illustrate that collective measurements with postselection can be
used to detect the nonlocality of a larger set of entangled states.

That postselection can be used to reveal such ``hidden
nonlocality'', was already shown in 1994 by
Popescu~\cite{S.Popescu:PRL:1995} using sequential measurements.
After that, Gisin~\cite{N.Gisin:PLA:1996} also demonstrated that
(without collective measurements) postselection itself in the form
of local filtering operations can be used to detect a larger set of
two-qubit entangled states. It is worth noting that an experimental demonstration of ``hidden
nonlocality'' has been reported  in~\cite{P.G.Kwiat:Nature:2001}.

In this paper, we will show that postselection is not necessary to
improve Bell inequality violation. In order to find such examples for
mixed states we have resorted to various
numerical approaches that are described in~\cite{Y.C.Liang:2005a} and
provide upper and lower bounds on the optimal violation of a given
Bell inequality by a given quantum state. The two algorithms described
in~\cite{Y.C.Liang:2005a} make use of convex optimization
techniques, specifically semidefinite
programs~\cite{L.Vandenberghe:SR:1996,S.Boyd:Book:2004}. The first,
henceforth  referred as the LB algorithm, is an algorithm that
can be used to determine, for a given quantum state $\rho$, a lower
bound of its maximal violation of a given Bell inequality. This can
be seen as an extension of the See-Saw iteration developed by Werner
and Wolf~\cite{R.F.Werner:QIC:2001} to Bell inequalities with more
than two outcomes. As with many other numerical optimization
techniques, the LB algorithm converges to a local maximum of the
global optimization problem, and hence, feeding the algorithm with
various random initial guesses is essential. Unless otherwise
stated, Bell inequality violations presented hereafter refer to the
best violation that we could find either analytically, or
numerically using this LB algorithm.

Complementarily, the other algorithm, which we shall refer as the UB
algorithm, is one that can be used to determine an upper bound on
the maximal violation of $\rho$ for a given Bell inequality. The
technique involves relaxing the complicated optimization over
measurements in the Bell experiment to a sequence of semidefinite
programs using techniques that have been developed in the general context of
non-linear optimization
theory~\cite{P.A.Parrilo:MP:2003,J.B.Lasserre:SJO:2001}  and applied in
quantum information theory in other
contexts~\cite{A.C.Doherty:,J.Eisert:PRA:2004}.  These methods provide global upper
bounds on the Bell inequality violation that can be accurately and
efficiently computed. The upper bounds obtained via this algorithm are
often not tight, but are sometimes
non-trivial~\cite{Y.C.Liang:2005a}. For ease of reference,
these upper bounds are marked where they appear with
$^\dag$. In the event that a
violation presented is known to be maximal (such as those computable
using the Horodecki's criterion~\cite{RPM.Horodecki:PLA:1995}), an
* will be attached.

This paper is organized as follows. In
Sec.~\ref{sec:measurements}, we present a measurement scheme which
we will use to determine the Bell-Clauser-Horne inequality violation
for any bipartite pure state. These measurements led to the largest
violation that we were able to find and may even be maximal. Then,
in Sec.~\ref{sec:pure-states}, we show that for bipartite pure
entangled  states, collective measurement can lead to a greater
violation of the Bell-CH inequality. The corresponding scenario for
mixed entangled states is analyzed in
Sec.~\ref{sec:mixed-states}. We then conclude with a summary of
results and some future avenues of research.

\section{Bell-CH-violation for Pure Two-Qudits}\label{sec:measurements}

In this section, we present a measurement scheme which gives rise to
the largest Bell-Clauser-Horne (henceforth abbreviated as Bell-CH)
inequality~\cite{CH:PRD:1974} violation that we have found for
arbitrary pure two-{\em qudit} states, i.e.~quantum states
describing a composite of two $d$-dimensional quantum subsystems.
We find using this inequality for probabilities rather than
correlations to be convenient for our purposes and the equivalence
between the Bell-CH inequality and the
Bell-Clauser-Horne-Shimony-Holt (henceforth abbreviated as
Bell-CHSH) inequality~\cite{CHSH:PRL:1969} in the ideal limit,
implies that if the conjectured measurement scheme is
optimal for the Bell-CH inequality, it will also give rise to the
maximal Bell-CHSH inequality violation for any pure two-qudit
state.

The Bell-CH inequality is meant for an experimental setup involving
two observers, Alice (A) and Bob (B). Each of these observers can
perform two alternative measurements, and each measurement gives
rise to two possible outcomes which we shall label by $\pm$. The
Bell-CH inequality is as follows~\cite{CH:PRD:1974}:
\begin{gather}
\Slhv=p_{AB}^{+-}(1,1)+p_{AB}^{+-}(1,2)+p_{AB}^{+-}(2,1)\nonumber\\
-p_{AB}^{+-}(2,2)-p_{A}^{+}(1)-p_{B}^{-}(1)\le 0,
\end{gather}
where $p_{AB}^{+-}(k,l)$ refers to the joint probability that
experimental outcome $+$ and $-$ are observed at A's
and B's site  respectively, given that Alice performs the $k^{\rm th}$ and
Bob performs the $l^{\rm th}$ measurement; the marginal
probabilities $p_A^+(k)$ and $p_B^-(l)$ are similarly defined. In
quantum mechanics, these probabilities are calculated according to
\begin{gather}
p_{AB}^{+-}(k,l)= \tr\left(\rho\,A_k^+\ten B_l^-\right)\nonumber\\
p_{A}^{+}(k)= \tr\left(\rho\,A_k^+\ten \unit_B\right),\quad
p_{B}^{-}(l)= \tr\left(\rho\,\unit_A\ten B_l^-\right),
\end{gather}
where we have denoted by $A^k_+$ the POVM element associated with
the ``+" outcome of Alice's $k^\text{th}$ measurement and $B^l_-$ the
POVM element associated with the ``-" outcome of Bob's $l^\text{th}$
measurement.

The maximal Bell inequality violation for a quantum state is
invariant under a local unitary transformation. As such, the maximal
Bell inequality violation for any bipartite pure quantum state is
identical to its maximal violation when written in the Schmidt
basis~\cite{E.Schmidt:MA:1906,S.M.Barnett:PLA:1991}. In this basis,
an arbitrary bipartite pure state in $d$-dimension, $\ket{\Psi_d}$
takes the form $\ket{\Psi_d}=\sum_{i=1}^d
c_i\ket{\varphi_i}_A\ket{\varphi_i}_B$, where
$\{\ket{\varphi_i}_A\}$ and $\{\ket{\varphi_i}_B\}$ are local
orthonormal bases of subsystem possessed by observer A and B
respectively, and $\{c_i\}_{i=1}^d$ are the  Schmidt coefficients of
$\ket{\Psi_d}$. Without loss of generality, we may also assume that
$c_1\ge c_2\ge\ldots\ge c_d\ge 0$. Then $\ket{\Psi}_d$ is entangled
if and only if $d>1$. Now, let's consider the following measurement
settings for Alice, which were first adopted
in~\cite{N.Gisin:PLA:1992},
\begin{gather}
A_1^{\pm}=\half\left[\unit_d\pm Z\right],\quad
A_2^{\pm}=\half\left[\unit_d\pm X\right],\nonumber\\
Z\equiv \oplus_{i=1}^{\lfloor d/2\rfloor}\sigma_z+\Pi,\quad X\equiv
\oplus_{i=1}^{\lfloor d/2\rfloor}\sigma_x+\Pi,\nonumber\\
\label{Eqn:Alice-POVM}\left[\Pi\right]_{ij}=0\quad\forall\quad i,j\neq d,\qquad
\left[\Pi\right]_{dd}=d\mod 2,
\end{gather}
where $\sigma_x$ and $\sigma_z$ are respectively the Pauli $x$ and $z$ matrices.

Notice, however, that the $\left\{B_l^\pm\right\}_{l=1}^2$ given
in~\cite{N.Gisin:PLA:1992} is not optimal. In fact, given the
measurements for Alice in Eqn.~(\ref{Eqn:Alice-POVM}), the optimization of
Bob's measurement
settings can be carried out explicitly~\cite{fn:Helstrom}. Using the
resulting analytic expression for Bob's optimal
POVM~\cite{Y.C.Liang:2005a}, the optimal expectation value of the Bell-CH
operator~\cite{S.L.Braunstein:PRL:1992} for
$\ket{\Psi_d}$ can be computed and we find
\begin{equation}\label{Eqn:Sqm:CH:pure}
\eBellCH_{\ket{\Psi_d}}=\half\sum_{n=1}^{\lfloor  d/2\rfloor} \sqrt{(c_{2n-1}^2+c_{2n}^2)^2+4
    c_{2n}^2 c_{2n-1}^2}+\frac{\gamma}{2}c_d^2-\half,
\end{equation}
where $\gamma\equiv d\mod 2$~\cite{fn:CH-CHSH}.

Effectively, this measurement scheme corresponds to first ordering
each party's local basis vectors $\{\ket{\varphi_i}\}_{i=1}^d$
according to their Schmidt coefficients, and grouping them pairwise in
descending order from the Schmidt vector with the largest Schmidt
coefficient. Physically, this can
be achieved by Alice and Bob each performing an appropriate local
unitary transformation. Each of their Hilbert space
can then be represented as a direct sum of 2-dimensional
subspaces, which can be regarded as a one-qubit space, plus a
1-dimensional subspace if $d$ is odd. The final step of the
measurement consists of performing the optimal
measurement~\cite{RPM.Horodecki:PLA:1995} in each of these two-qubit
spaces as if the other spaces did not exist.

From here, it is easy to see that if we have a maximally entangled
state, i.e. $\ket{\Psi_d}_{\rm
ME}=\frac{1}{\sqrt{d}}\sum_{i=1}^{d}\ket{\varphi_i}_A\ket{\varphi_i}_B$,
then \eqref{Eqn:Sqm:CH:pure} gives
\begin{equation}\label{Eqn:S:ME}
\eBellCH_{\ket{\Psi_d}_{\rm ME}}=\left\{\begin{array}{r@{\quad \quad}}
 \frac{1}{\sqrt{2}}-\half^*:
 d\,\rm{even} \\
\frac{\sqrt{2}(d-1)+1}{2d}-\half: d~\rm{odd} \\  \end{array}\right. .
\end{equation}
Under this measurement scheme, the  Bell-CH inequality
violation for a maximally entangled state with even $d$ is
thus the maximum allowed by Cirelson's
bound~\cite{B.S.Cirelson:LMP:1980} whereas that of maximally
entangled state with odd $d$ is
not.

How good is the measurement scheme \eqref{Eqn:Alice-POVM}? It is
constructed so that for
two-qubits, i.e. when $d=2$, \eqref{Eqn:Sqm:CH:pure} gives the same
violation found in~\cite{N.Gisin:PLA:1991,N.Gisin:PLA:1992}, and is
the maximal  violation determined by Horodecki
et~al.~\cite{RPM.Horodecki:PLA:1995}. The measurement given by
\eqref{Eqn:Alice-POVM} is hence optimal for two-qubit states. For
higher dimensional quantum systems, we have looked at randomly
generated pure two-qudit states ($d=3,\ldots,10$) with their
(unnormalized) Schmidt coefficients uniformly chosen at random from
the interval $(0,1)$. For all of the 20,000 states generated for
each $d$, we found that with \eqref{Eqn:Alice-POVM} as the initial
measurement setting, the (iterative) LB algorithm never gives a
$\eBellCH_{\ket{\Psi_d}}$ that is different from
\eqref{Eqn:Sqm:CH:pure} by more than $10^{-15}$, thus indicating
that \eqref{Eqn:Sqm:CH:pure} is, at least, a local maximum of the
optimization problem.

Furthermore, for another 8,000 randomly generated pure two-qudit
states, 1,000 each for $d=3,\ldots,10$, an extensive  numerical search
using more than $4.6\times 10^6$ random initial measurement guesses
have not led to a single instance where $\eBellCH_{\ket{\Psi_d}}$ is
higher than that given in
\eqref{Eqn:Sqm:CH:pure}~\cite{fn:close-to-optimal}. These numerical
results suggest that the measurement scheme given by
\eqref{Eqn:Alice-POVM} may be the optimal measurement that maximizes
the Bell-CH inequality violation for arbitrary pure two-qudit
states.

\section{Multiple Copies of Pure States}\label{sec:pure-states}
Let's now look into the problem of whether nonlocal correlations can
be enhanced by performing collective measurements on $N>1$ copies of
an entangled quantum state~\cite{fn:SN>SM}. As our first example of
nonlocality  enhancement, consider again those maximally entangled
states residing in Hilbert space with odd $d$. It is well known
their maximal Bell-CH/ Bell-CHSH inequality violation cannot
saturate Cirelson's bound~\cite{S.Popescu:PLA:1992b}. In fact, their
best known Bell-CH inequality violation~\cite{N.Gisin:PLA:1992} is
that given in~\eqref{Eqn:S:ME}. By combining $N$ copies of these
quantum states, it is readily seen that we effectively end up with
another maximally entangled state of $d^N$-dimension. It then
follows from \eqref{Eqn:S:ME} that their Bell-CH violation under
collective measurements  increases monotonically with the  number of
copies $N$ (see also Table~\ref{tbl:MaximalViolation}, column 3 and
7). In fact, it can be easily shown that this violation approaches
asymptotically the Cirelson's bound~\cite{B.S.Cirelson:LMP:1980} in
the limit of large $N$. Therefore, if the maximal violation of these
quantum states is given by \eqref{Eqn:S:ME}, collective measurements
can already give better Bell-CH violation with $N=2$. Even if the
maximal violation is not given by \eqref{Eqn:S:ME}, it can be seen,
by comparing the upper bound of the single-copy violation from the UB
algorithm and the
lower bound of the $N$-copy violation, from
Table~\ref{tbl:MaximalViolation} that for $d=3$ and $d=5$, a Bell-CH
violation better than the maximal single-copy violation can always
be obtained when $N$ is sufficiently large.

Such an enhancement is even more pronounced in the case of
non-maximally entangled states. In particular, for $N$ copies of a
(non-maximally entangled) two-qubit state written in the Schmidt
basis,
\begin{equation}\label{Eqn:NME-2-qubit}
\ket{\Psi_2}^{\otimes  N}=\left(\cos\phi\ket{00}+\sin\phi\ket{11}\right)^{\otimes  N},
\end{equation}
where $0<\phi\le\frac{\pi}{4}$~\cite{fn:phi}. The Bell-CH violation given by \eqref{Eqn:Sqm:CH:pure} is
\begin{equation}\label{eqn:S:NME}
\eBellCH_{\ket{\Psi_2}}=\frac{p}{\sqrt{2}}+\frac{1-p}{2}\sqrt{1+\sin^22\phi}-\half,
\end{equation}
where
\begin{equation*}
p=1-\half\cos^{2(N-1)}\phi\sum_{m=0}^{N-1}\tan^{2m}\phi\left[1-(-1)^{\frac{(N-1)!}{m!(N-1-m)!}}\right],
\end{equation*}
is the total probability of finding $\ket{\Psi_2}^{\ten N}$  in one
of the {\em perfectly correlated 2-dimensional subspaces} (i.e.~a
subspace with $c_{2n-1}=c_{2n}$) upon reordering  of the Schmidt
coefficients in descending order.

\begin{figure}[h!]
\includegraphics[width=9.8cm,height=7.6cm]{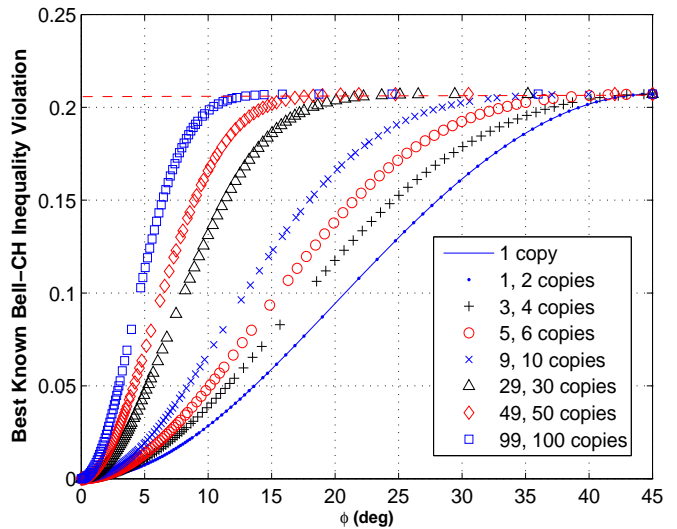}
\caption{\label{Fig:S:NME:Qubit} (Color online) Best known Bell-CH inequality
violation of pure two-qubit states obtained
from~\eqref{Eqn:Alice-POVM}, plotted as  a function of $\phi$, which
gives a primitive measure of entanglement; $\phi=0$ for bipartite
pure product state and $\phi=45^o$ for bipartite maximally entangled
state. The curves from right to left represent increasing numbers of
copies. The dotted horizontal line at $\frac{1}{\sqrt{2}}-\half$ is
the maximal possible violation of Bell-CH inequality; correlations
allowed by local realistic theories have values less than or equal
to zero. The solid line is the maximal Bell-CH inequality violation of
$\ket{\Psi_2}$ determined using the Horodecki's criterion~\cite{RPM.Horodecki:PLA:1995}.}
\end{figure}

It is interesting to note that for these two-qubit states, their
Bell-CH inequality violation for $N=2k-1$ copies, and $N=2k$ copies
are identical~\cite{fn:degenerate} for all $k\ge 1$, as illustrated
in the second column of Table~\ref{tbl:MaximalViolation} and in
Fig.~\ref{Fig:S:NME:Qubit}. This feature, however, does not seem to
generalize to higher dimensions.

Like the odd-dimensional maximally entangled state, the violation of the
Bell-CH inequality for {\em any} pure two-qubit entangled states, as
given by \eqref{Eqn:Sqm:CH:pure},  increases asymptotically towards
the Cirelson bound~\cite{B.S.Cirelson:LMP:1980} with the number of
copies $N$, as can be seen in Fig.~\ref{Fig:S:NME:Qubit}. A direct
implication of this is that, with a sufficiently large number of
copies, the nonlocality present in {\em any} weakly entangled pure
two-qubit states is of no noticeable difference from that in a
maximally entangled two-qubit state.

\begin{table}[!h]
\caption{\label{tbl:MaximalViolation} Best known Bell-CH inequality
violation for some bipartite pure entangled states, obtained from
\eqref{Eqn:Alice-POVM} with and without  collective measurements.
Also included below is the upper bound of $\eBellCH_{\ket{\Psi}}$
obtained from the UB algorithm. The first column of the table gives the
number of copies $N$ involved in the measurements. Each quantum
state is labeled by their non-zero Schmidt coefficients, which are
separated by : in the subscripts attached to the ket vectors;
e.g.~$\ket{\Psi}_{1:2:3:3}$  is the state with unnormalized Schmidt
coefficients $\{c_i\}_{i=1}^4=\{1,2,3,3\}$. For each quantum state
there is a box around the entry corresponding to the smallest $N$ such
that the lower bound of $\eBellCH_{\ket{\Psi}}$ on the maximal violation
exceeds the single-copy upper bound coming from the UB algorithm.}
\begin{ruledtabular}
\begin{tabular}{r|cccccc}
$N$ & $\ket{\Psi_{2:1}}$ & $\ket{\Psi_{1:1:1}}$
    & $\ket{\Psi_{1:2:3}}$ & $\ket{\Psi_{1:2:3:4}}$ &
    $\ket{\Psi_{1:2:3:3}}$ & $\ket{\Psi_{1:1:1:1:1}}$ \\ \hline
  &Lower & Bound& & & & \\
\hline
  1 & 0.14031* & 0.13807 & 0.16756 & 0.18431 & 0.19259 & 0.16569\\
  2 & 0.14031 & 0.18409 & 0.18307 & 0.19624 & 0.20516 & 0.19882\\
  3 & \boxed{0.16169} & \boxed{0.19944} & 0.19451 & 0.20275 & 0.20685 & 0.20545\\
  4 & 0.16169 & 0.20455 & \boxed{0.19642} & 0.20388 & 0.20706 & \boxed{0.20678}\\
  5 & 0.17964 & 0.20625 & 0.20254 & 0.20596 & 0.20710 & 0.20704\\
 10 & 0.19590 & 0.20710 & 0.20643 & 0.20704 & 0.20711 & 0.20711 \\
 \hline
  &Upper & Bound& & & & \\
\hline
  1 & 0.14031* & 0.18409$^\dag$ & 0.19624$^\dag$ & 0.20711$^\dag$ & 0.20711$^\dag$ & 0.20569$^\dag$ \\
\end{tabular}
\end{ruledtabular}
\end{table}

Similarly, if we consider $N$ copies of pure two-qutrit entangled states
written in the Schmidt form,
\begin{equation}
\ket{\Psi_3}^{\otimes
N}=\left(\cos\phi\ket{00}+\sin\phi\cos\theta\ket{11}+
\sin\phi\sin\theta\ket{22}\right)^{\otimes N},
\end{equation}
where $0<\phi\le\frac{\pi}{4}$, $0<\theta\le\frac{\pi}{4}$, it can
be verified that their Bell-CH inequality violation, as given by
\eqref{Eqn:Sqm:CH:pure}, also increases steadily with the number of
copies. Thus, if \eqref{Eqn:Sqm:CH:pure} gives the maximal Bell-CH
violation for pure two-qutrit states, better Bell inequality
violation can also be attained by collective measurements using two
copies of these quantum states. The explicit value of the violation
can be found in  column 3 and 4 of Table~\ref{tbl:MaximalViolation}
for two specific two-qutrit states. As above, even if the maximal
Bell-CH violation is not given by \eqref{Eqn:Sqm:CH:pure},
collective measurements with \eqref{Eqn:Alice-POVM} can definitely give a
violation that is better than the maximal-single-copy ones as a result
of the bound coming from the UB algorithm for a single copy (see
Table~\ref{tbl:MaximalViolation}). Corresponding examples for pure
bipartite four-dimensional and five-dimensional quantum states can
also be found in the table.

Some intuition for the way in which better Bell-CH inequality violation may be
obtained with collective measurements and the measurement scheme
\eqref{Eqn:Alice-POVM} is that the reordering of
subspaces prior to the measurements \eqref{Eqn:Alice-POVM} generally
increases the total probability of finding 2-dimensional subspaces
with $c_{2n}=c_{2n-1}$, while ensuring that the remaining
2-dimensional subspaces are at least as correlated as any of the
corresponding single-copy 2-dimensional subspaces. The measurement
then effectively projects onto each of these subspaces (with Alice and
Bob being guaranteed to obtain the same result) and then performs the
optimal measurement on the resulting shared two-qubit state. Since the optimal
measurements in each of these perfectly correlated 2-dimensional
subspaces give the maximal Bell-CH inequality violation, while the
same measurements in the remaining 2-dimensional subspaces give as
much violation as the single-copy violation, the multiple-copy
violation is thus generally greater than that of a single copy.

As one may have noticed, our measurement protocol bears some
resemblance with the entanglement concentration protocol developed
by  Bennett et~al.~\cite{C.H.Bennett:PRA:1996}. In entanglement
concentration Alice and Bob
make slightly different projections onto subspaces that are spanned by
all those ket vectors sharing the same Schmidt
coefficients thus obtaining a maximally entangled state in a bipartite
system of some dimension. One can also obtain improved Bell inequality
violations by
adopting their protocol and first projecting Alice's Hilbert space
into one of the perfectly correlated subspaces  and performing the
best known measurements for a Bell inequality violation in each of
these (not necessary 2-dimensional) subspaces. We have compared the
Bell-CH inequality violation of an arbitrary pure two-qubit state
derived from each of these protocols and found that the violation
obtained using our protocol always outperforms the other. The
difference, nevertheless, diminishes as $N\to\infty$. This
observation provides another consistency check of the optimality
of~\eqref{Eqn:Sqm:CH:pure}.

\section{Multiple Copies of Mixed States}\label{sec:mixed-states}
The impressive enhancement in a pure state Bell-CH inequality violation naturally leads us to
ask if the same conclusion can be drawn for mixed entangled states.
The possibility of obtaining better Bell inequality violation with
collective measurements, however, does not seem to generalize to all
entangled states.

\begin{figure}[h!]
\centering\rule{0pt}{4pt}\par
\includegraphics[width=9.8cm,height=7.6cm]{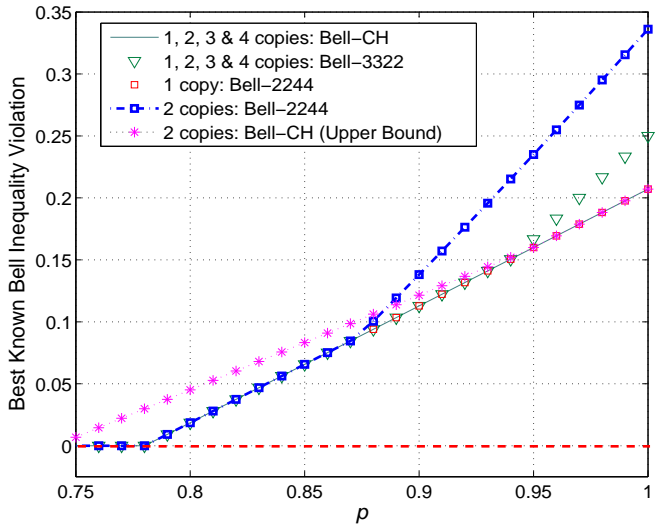}
\caption{\label{Fig:Werner} (Color online) Best known expectation value of the
Bell-CH, Bell-3322 and Bell-2244 operators with respect to the
2-dimensional Werner states; $p$ represents the overlap with
a singlet state. Also included is the upper bound on the
maximal $\eBellCH_{\rho_{W}^{\ten 2}}$ obtained from the UB
algorithm~\cite{Y.C.Liang:2005a}.}
\end{figure}

Our first counterexample comes from the 2-dimensional Werner
state~\cite{R.F.Werner:PRA:1989}, which can seen as a mixture of
singlet state and the maximally mixed state,
\begin{equation}
\rho_w=(1-p)\frac{\unit_4}{3}+\frac{4p-1}{3}\ketbra{\Psi^-},
\end{equation}
where $p$ is the probability of finding a singlet state in this
mixture. This state is entangled for $p>\half$ and violates the
Bell-CH inequality if and only if~\cite{RPM.Horodecki:PLA:1995}
$p>p_w\equiv\frac{1}{4}\left(\frac{3}{\sqrt{2}}+1\right)\simeq 0.7803$.
Using the LB algorithm~\cite{Y.C.Liang:2005a}, we have searched for
the maximal violation of $\rho_w$ with $p>p_w$ for $N\le 4$ copies
but no increase in the maximal violation of Bell-CH inequality has
ever been observed  (see Fig.~\ref{Fig:Werner}). In fact, by using
the UB algorithm~\cite{Y.C.Liang:2005a}, we find that for two copies
of some Bell-CH violating Werner states, their maximal Bell-CH
inequality violation are identical to the corresponding single-copy
violation within numerical precision of $10^{-12}$. This strongly
suggests that for some Werner states the maximal Bell-CH inequality
violation does not depend on  the number of copies $N$.

\begin{figure}[h!]
\centering\rule{0pt}{4pt}\par
\includegraphics[width=9.8cm,height=7.6cm]{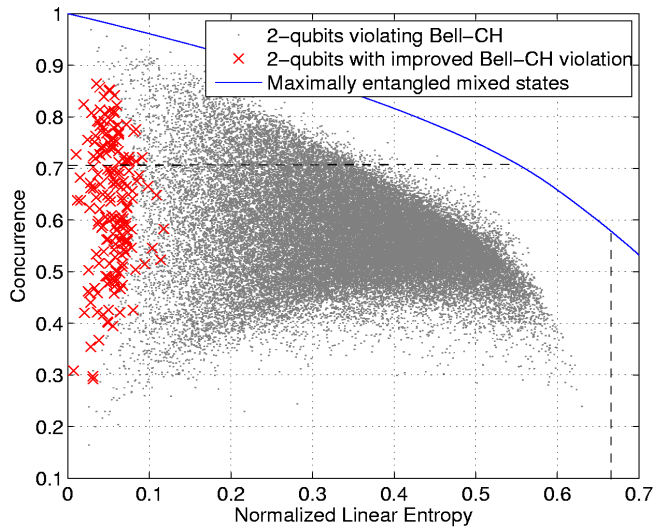}
\caption{\label{Fig:CS-CH-Enhancement} (Color online) Distribution of two-qubit
states sampled for improved Bell-CH violation by collective
measurements. The maximally entangled mixed states (MEMS), which
demarcate the boundary of the set of density matrices on this
concurrence-entropy
plane~\cite{W.J.Munro:PRA:2001,T-C.Wei:PRA:2003}, are represented by
the solid line. Note that as a result of the chosen distribution over
mixed states~\cite{fn:random states} this region is not well
sampled. The region bounded by
the solid line and the
horizontal dashed line (with concurrence equal to $1/\sqrt{2}$)
only contain two-qubit states that {\em violate} the Bell-CH
inequality~\cite{L.Derkacz:PRA:2005}; the region bounded by the
solid line and the vertical dashed line (with normalized linear
entropy equal to $2/3$) only contain states that {\em do
not violate} the Bell-CH
inequality~\cite{E.Santos:PRA:2004,L.Derkacz:PRA:2005}. Two-qubit
states found to give better 3-copy Bell-CH violation are marked with
red crosses.}
\end{figure}

There are, nevertheless, some two-qubit states whose maximal Bell-CH
inequality violation for $N=3$ is higher than the corresponding
single-copy violation. In contrast to the pure state scenario, the
set of mixed two-qubit states seems to be dominated by those whose
3-copy Bell-CH inequality violation is not enhanced. In fact, among
50,000  randomly generated Bell-CH violating two-qubit
states~\cite{fn:random states},  only about $0.38\%$ of them were
found to have their 3-copy Bell-CH inequality violation greater than
their maximal single-copy violation. Moreover, as can be seen in
Fig.~\ref{Fig:CS-CH-Enhancement}, they are all clustered at regions
with relatively low linear entropy.

\begin{figure}[h!]
\centering\rule{0pt}{4pt}\par
\includegraphics[width=9.8cm,height=7.6cm]{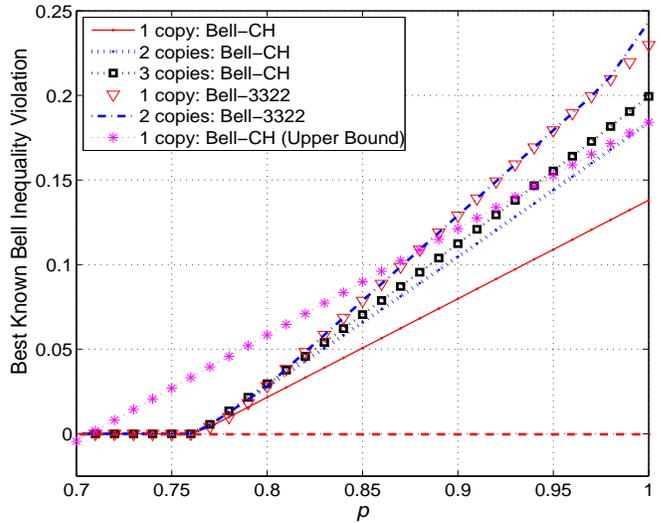}
\caption{\label{Fig:MEQt} (Color online) Best known expectation value of the
Bell-CH and Bell-3322 operators with respect to the 3-dimensional
isotropic states; $p$ is the fraction of maximally entangled
two-qutrit state in the mixture. Also included is the upper bound on the
maximal $\eBellCH_{\rho_{I_3}}$ obtained from the UB
algorithm~\cite{Y.C.Liang:2005a}.}
\end{figure}

As with the pure state scenario, an enhancement of nonlocal
correlations in the Bell-CH setting seems to be more prevalent in
higher dimensional quantum systems. In particular, for all of the
3-dimensional isotropic states~\cite{MPR.Horodecki:0109124}
\begin{equation}
\rho_{I_3}=p\,\ket{\Psi_3}_\text{ME}\bra{\Psi_3}+(1-p)\frac{\unit_9}{9}
\end{equation}
that were found to violate the Bell-CH inequality, numerical results
obtained from the LB algorithm suggest that the maximal violation
increases steadily with the number of copies. Further results
obtained using the UB algorithm show that with $N=3$, some of the
Bell-CH violating $\rho_{I_3}$ definitely give better Bell-CH
violation with collective measurements. The results are summarized in
Fig.~\ref{Fig:MEQt}.

Yet another question that one can ask is how much does the
enhancement of nonlocal correlations depend on the choice of Bell
inequality.  To address this question, we have also studied the
enhancement of nonlocal correlations with respect to other Bell
inequalities for probabilities, in particular the Bell-3322
inequality, the Bell-2233 inequality and the Bell-2244
inequality~\cite{D.Collins:JPA:2004,fn:Enumeration}. For these Bell
inequalities, we find that the possibility of enhancing nonlocal
correlations does seem to depend on both the number of alternative
settings and the number of possible outcomes involved in a Bell
experiment. The dependence on the number of outcomes is particularly
prominent in the case of Werner states, where a large range of
Bell-2244-inequality-violating Werner states seem to achieve a
higher two-copy violation, even though their maximal Bell-CH
inequality violation apparently remains unchanged up to $N=4$
(Fig.~\ref{Fig:Werner}).

The dependence on the number of alternative settings can be seen in
the best known violation of $\rho_{I_3}$ with respect to the Bell-CH
inequality and the Bell-3322 inequality (Fig.~\ref{Fig:MEQt}). In
particular, when the number of alternative settings is increased
from 2 (in the case of Bell-CH inequality) to 3 (in the case of
Bell-3322 inequality), the range of states whereby collective
measurements were found to improve the Bell inequality
violation is drastically {\em reduced}.\\

\section{Conclusion}
In this paper, we have focused on bipartite entangled systems  and
considered the enhancement of nonlocal correlations by collective
measurements without postselection. This amounts to allowing an experiment in which
a local unitary is applied to a number of copies of the state $\rho$
prior to the Bell inequality experiment.

We find that the Bell-CH inequality violation of all bipartite pure
entangled states, can be enhanced by allowing collective measurements
even without postselection. For mixed entangled states, however,
explicit examples (Werner states) have been presented to
demonstrate that there may be entangled states whose nonlocal
correlations cannot be enhanced in any Bell-CH experiments. In fact,
the set of mixed two-qubit states whose Bell-CH violation can be increased
with collective measurements seems to be relatively small.

We have also done some preliminary studies on how the usefulness of
collective measurements depends
on the choice of Bell inequality and on the dimension of the
subsystem. Our data at the moment are consistent with the hypothesis
that the usefulness of collective measurements in Bell inequality
experiments increases with the Hilbert space dimension and with the number
of measurement outcomes allowed by Bell inequality. On the other hand
as the number of measurement settings allowed by the Bell inequality
increases the advantage provided by collective measurements seems to
diminish. However, note that we have not really performed the
systematic study required to establish such trends, if they exist, due
to the great numerical effort that would be required.
Given these observations, it does seem that postselection is a lot
more powerful than collective measurements on their own in increasing
Bell inequality violation.

An immediate question that follows from the present work is what is
the class of quantum states whereby collective measurements can
increase their Bell inequality violation? One motivation for studying our
problem is to understand better the set of quantum states that violate
a Bell inequality and are thus inconsistent with local realism.  It has
been known for a long time that this set is a strict subset of the
entangled states if projective~\cite{R.F.Werner:PRA:1989} or even generalized
measurements~\cite{J.Barrett:PRA:2002} on single copies of a system are permitted. One
might wonder whether collective measurements without postselection allow
us to violate Bell inequalities for a larger set of states. However we do
not know of examples where a state that does not violate a given Bell inequality
becomes violating under collective measurements when no
postselection is allowed. Moreover, for mixed states, the set of
states whose violations increase when collective measurements are
allowed appears to be rather restricted. This is consistent with the
recent work by
Masanes~\cite{Ll.Masanes:eprint:0512153} which suggests that the set of
states that violates a given Bell inequality under collective
measurements without postselection is a subset of all distillable
states.

\begin{acknowledgments}
This work was supported by the Australian Research Council. We thank
Ben Toner for stimulating discussions, and Henry Haselgrove and
Eric Cavalcanti for interesting comments. In addition, Y.C.~Liang
gratefully acknowledges the kind hospitality of the quantum information
technology group of NUS, where part of this work was completed;
helpful suggestions from Reinhard Werner, Jing Ling Chen, and Meng
Khoon Tey are particularly appreciated.
\end{acknowledgments}


\begin{thebibliography}{99}

\bibitem{J.S.Bell:1964} J.~S. Bell, Physics, {\bf 1}, 195 (1964).

\bibitem{CHSH:PRL:1969} J.~F.~Clauser, M.~A.~Horne, A.~Shimony and
R.~Holt, \prl {\bf 23}, 880 (1969).

\bibitem{CH:PRD:1974} J.~F.~Clauser, and M.~A.~Horne, \prd
 {\bf 10}, 526 (1974).

\bibitem{N.Gisin:PLA:1991} N.~Gisin, Phys.~Lett.~A {\bf 154}, 201 (1991).

\bibitem{N.Gisin:PLA:1992} N.~Gisin and A.~Peres, Phys.~Lett.~A {\bf 162}, 15 (1992).

\bibitem{S.Popescu:PLA:1992} S.~Popescu and D.~Rohrlich, \pl A {\bf
166}, 293 (1992).

\bibitem{A.Aspect:Nature:1999} A.~Aspect, Nature {\bf 398}, 189 (1999).

\bibitem{M.A.Rowe:Nature:1999} M.~A.~Rowe, D.~Kielpinski, V.~Meyer,
  C.~A.~Sackett, W.~A,~Itano, C.~Monroe, and D.~J.~Wineland, Nature {\bf 409}, 791 (1999).

\bibitem{R.F.Werner:QIC:2001} R.~F.~Werner and M.~M.~Wolf, Quantum
  Information and Computation {\bf 1}, 1 (2001).

\bibitem{A.Peres:PRA:1996} A.~Peres, \pra {\bf 54}, 2685 (1996).

\bibitem{fn:Peres} The postselection in Peres' scheme is stronger
than that in realistic Bell inequality experiments where detector
inefficiencies require a postselection on events where both detectors
fired. In such a case the failure probability is independent of the
quantum state.

\bibitem{S.Popescu:PRL:1995} S.~Popescu, \prl {\bf 74}, 2619 (1995).

\bibitem{N.Gisin:PLA:1996} N.~Gisin, \pl A {\bf 210}, 151 (1996).

\bibitem{P.G.Kwiat:Nature:2001} P.~G.~Kwiat, S.~Barraza-Lopez,
  A.~Stefanov, and N.~Gisin, Nature {\bf 409}, 1014 (2001).

\bibitem{Y.C.Liang:2005a} Y.~C.~Liang and A.~C.~Doherty, eprint
quant-ph/0608128.

\bibitem{L.Vandenberghe:SR:1996} L.~Vandenberghe and S.~Boyd, SIAM Review {\bf
  38}, 49 (1996).

\bibitem{S.Boyd:Book:2004} S.~Boyd and L.~Vandenberghe, {\em Convex
  Optimization} (New York: Cambridge, 2004).

\bibitem{P.A.Parrilo:MP:2003} P.~A.~Parrilo, Math.~Program {\bf 96}, 293 (2003).

\bibitem{J.B.Lasserre:SJO:2001} J.~B.~Lasserre, SIAM J. Optim. {\bf 11}, 796 (2001).

\bibitem{A.C.Doherty:} A.~C.~Doherty, P.~A.~Parrilo, and
  F.~M.~Spedalieri, \pra {\bf 69}, 022308 (2004).

\bibitem{J.Eisert:PRA:2004} J.~Eisert, P.~Hyllus, O.~G\"uhne, and
  M. Curty, \pra {\bf 70}, 062317 (2004).

\bibitem{RPM.Horodecki:PLA:1995} R.~Horodecki, P.~Horodecki, M.~Horodecki,  \pl A  {\bf 200}, 340 (1995).

\bibitem{E.Schmidt:MA:1906} E. Schmidt, Math. Ann. {\bf 63}, 433 (1906).

\bibitem{S.M.Barnett:PLA:1991} S.~M.~Barnett and S.~J.~D.~Phoenix,
  Phys.~Lett.~A {\bf 167}, 233 (1992).

\bibitem{fn:Helstrom} The calculation is essentially the same as
that which shows that the Helstrom measurement~\cite{C.W.Helstrom:Book:1976} is optimal for
distinguishing two quantum states.

\bibitem{C.W.Helstrom:Book:1976} C.~W.~Helstrom, {\em Quantum
  detection and estimation theory} (Academic: NY, 1976).

\bibitem{S.L.Braunstein:PRL:1992} S.~L.~Braunstein, A.~Mann, and
  M.~Revzen, \prl {\bf 68}, 3259 (1992).

\bibitem{fn:CH-CHSH} The corresponding expectation value of the
Bell-CHSH operator can be obtained via $\eBellCHSH_{\ket{\Psi_d}}=4\left(\eBellCH_{\ket{\Psi_d}}+\half\right)$.

\bibitem{B.S.Cirelson:LMP:1980} B.~S.~Cirel'son,
  Lett. Math. Phys. {\bf 4}, 93 (1980).

\bibitem{fn:close-to-optimal} It is worth noting that among the
1,000  random pure states generated for each $d$, there are always
some  whose best Bell-CH inequality violation found differs from
 \eqref{Eqn:Sqm:CH:pure} by no more than $10^{-10}$.

\bibitem{fn:SN>SM} Notice that the maximal Bell
inequality violation for $N>M$ copies of a quantum system is never
less than that involving only $M$ copies. This follows from the fact
that the maximal $M$-copy violation can always be recovered in the
$N$-copy scenario by performing the $M$-copy-optimal-measurement on
$M$ of the $N$ copies, while leaving the remaining $N-M$ copies
untouched.

\bibitem{S.Popescu:PLA:1992b} S.~Popescu and D.~Rohrlich, \pl A {\bf
169}, 411 (1992).

\bibitem{fn:phi} For $\frac{\pi}{4}<\phi\le\frac{\pi}{2}$, we just have to redefine
$\phi$ as $\frac{\pi}{2}-\phi$ and all the subsequent results
follow.

\bibitem{fn:degenerate} This can be rigorously shown using
combinatoric arguments (private communication, Henry Haselgrove).

\bibitem{C.H.Bennett:PRA:1996} C.~H.~Bennett, H.~J.~Bernstein,
  S.~Popescu and B.~Schumacher, \pra {\bf 53}, 2046 (1996).

\bibitem{R.F.Werner:PRA:1989} R.~F.~Werner, \pra {\bf 40}, 4277 (1989).

\bibitem{fn:random states} We follow the algorithm presented
in~\cite{K.Zyczkowski:PRA:1998} to generate random two-qubit states.
In particular, the eigenvalues $\{\lambda_i\}_{i=1}^4$ of the
quantum states were chosen from a uniform distribution on the
4-simplex defined by $\sum_i \lambda_i=1$.

\bibitem{K.Zyczkowski:PRA:1998} K.~\.Zyczkowski, P.~Horodecki,
A.~Sanpera, and M.~Lewenstein, \pra {\bf 58}, 883 (1998).

\bibitem{W.J.Munro:PRA:2001} W.~J.~Munro, D.~F.~V.~James, A.~G.~White,
  and P.~G.~Kwiat, \pra {\bf 64}, 030302(R) (2001).

\bibitem{T-C.Wei:PRA:2003} T-C.~Wei, K.~Nemoto, P.~M.~Goldbart,
  P.~G.~Kwiat, W.~J.~Munro, and F.~Verstraete, \pra {\bf 67}, 022110
  (2003).

\bibitem{E.Santos:PRA:2004} E.~Santos, \pra {\bf 70}, 059901(E) (2004).

\bibitem{L.Derkacz:PRA:2005} {\L}.~Derkacz and L.~Jak\'obczyk, \pra {\bf
  72}, 042321 (2005).

\bibitem{MPR.Horodecki:0109124} M.~Horodecki, P.~Horodecki and
  R.~Horodecki, eprint quant-ph/0109124.

\bibitem{D.Collins:JPA:2004} D.~Collins and N.~Gisin, J.~Phys.~A {\bf 37}, 1775  (2004).

\bibitem{fn:Enumeration} We are adopting the notation
  in~\cite{D.Collins:JPA:2004} to enumerate the various tight Bell
inequalities for probabilities; a Bell-$m_Am_Bn_An_B$ inequality is
a Bell inequality for probability that involves two observers A and
B, where they can respectively perform one of the $m_A$ and $m_B$
alternative measurements, with each measurements yielding one of the
$n_A$ and $n_B$ possible outcomes.

\bibitem{J.Barrett:PRA:2002} J.~Barrett, \pra {\bf 65}, 042302 (2002).

\bibitem{Ll.Masanes:eprint:0512153} Ll.~Masanes, \prl {\bf 97}, 050503 (2006).

\end{thebibliography}
\end{document}